\documentclass[a4paper,11pt]{article}
\pdfoutput=1 

\usepackage{jinstpub} 
\usepackage[utf8]{inputenc}
\usepackage[binary-units=true]{siunitx}
\usepackage{tikz}
\usepackage[acronym]{glossaries}

\usetikzlibrary{shapes,arrows,positioning}
\tikzset{
  pics/carc/.style args={#1:#2:#3}{
    code={
      \draw[pic actions] (#1:#3) arc(#1:#2:#3);
    }
  }
}

\DeclareSIUnit\lightspeed{c}

\newcommand{\pythia}{\textsc{Pythia8}}
\newcommand{\geant}{\textsc{Geant4}}

\glsdisablehyper
\newacronym{HL-LHC}{HL-LHC}{High Luminosity upgrade of the LHC}
\newacronym{LHC}{LHC}{Large Hadron Collider}
\newacronym{AM}{AM}{Associative Memory}
\newacronym[plural=RoIs,longplural={Regions of Interest}]{RoI}{RoI}{Region of Interest}
\newacronym[plural=FPGAs,longplural={Field Programmable Gate Arrays}]{FPGA}{FPGA}{Field Programmable Gate Array}
\newacronym[plural=ASICs,longplural={Application-Specific Integrated Circuits}]{ASIC}{ASIC}{Application-Specific Integrated Circuit}
\newacronym{EF}{EF}{Event Filter}
\newacronym{ITk}{ITk}{Inner Tracker}
\newacronym{L0A}{L0A}{Level-0 Accept}
\newacronym{L1A}{L1A}{Level-1 Accept}
\newacronym{R3}{R3}{Regional Readout Request}
\newacronym{SM}{SM}{Standard Model}
\newacronym{BSM}{BSM}{Beyond Standard Model}
\newacronym{L1Track}{L1Track}{Level-1 Track trigger}
\newacronym{pt}{\ensuremath{p_\text{T}}}{transverse momentum}
\newacronym{eta}{\ensuremath{\eta}}{pseudorapidity}

\title{\boldmath Comparison of two hardware-based hit filtering methods for trackers in high-pileup environments.}

\author[a,b]{J. Gradin,}
\author[a]{M. Mårtensson}
\author[a]{and R. Brenner}

\affiliation[a]{Uppsala Universitet, Lägerhyddsvägen 1, 752 37 Uppsala, Sweden}
\affiliation[b]{Universite Grenoble Alpes, LPSC, 53 Avenue des Martyrs 38026 Grenoble cedex, France}

\emailAdd{joakim.gradin@physics.uu.se}

\abstract{As experiments in high energy physics aims to measure increasingly rare processes, the experiments continually strive to increase the expected signal yields. In the case of the High Luminosity upgrade of the LHC, the luminosity is raised by increasing the number of simultaneous proton-proton interactions, so-called pile-up. This increases the expected yields of signal and background processes alike.
  The signal is embedded in a large background of processes that mimic that of signal events. It is therefore imperative for the experiments to develop new triggering methods to effectively distinguish the interesting events from the background.

  We present a comparison of two methods for filtering detector hits to be used for triggering on particle tracks: one based on a pattern matching technique using Associative Memory (AM) chips and the other based on the Hough transform. Their efficiency and hit rejection are evaluated for proton-proton collisions with varying amounts of pile-up using a simulation of a generic silicon tracking detector. It is found that, while both methods are feasible options for an efficient track trigger, the AM based pattern matching produces a lower number of hit combinations with respect to the Hough transform whilst keeping more of the true signal hits. We also present the effect on the two methods when increasing the amount of support material in the detector and introducing inefficiencies by deactivating detector modules. The increased support material has negligable effects on the efficiency for both methods, while dropping \SI{5}{\%} (\SI{10}{\%}) of the available modules decreases the efficiency to about \SI{95}{\%} (\SI{87}{\%}) for both methods, irrespectively of the amount of pile-up.
}


\newcommand{\figures}{./fig}

\begin{document}
\maketitle
\flushbottom

\section{Introduction}
\label{sec:intro}

At the \gls{LHC}, experiments such as ATLAS~\cite{ATLAS} and CMS~\cite{CMS} are analyzing proton-proton interactions to study the nature of matter. The rare phenomena that are searched for are hidden in an enormous background from well know physics interactions. To increase sensitivity to interesting phenomena, the instantaneous luminosity in the \gls{LHC} will be increased~\cite{hl-lhc} to \SI{5e34}{\per\cm\squared\per\s}, which is 5 times higher than the design value of the current \gls{LHC}. The drawback of the higher luminosity is that it will also increase in the number of interactions per bunch crossing, referred to as \emph{pile-up}, by an equal amount. This presents new technical challenges to the experiments.

The sheer amount of data generated by the collisions at the \gls{HL-LHC} makes it impossible to read out and store all events. That is why experiments such as ATLAS and CMS use a \emph{trigger system} to select the most interesting events. It is particularly important to be able to trigger on high \gls{pt} particles, since they provide a clean signature for a variety of interesting processes in and beyond the Standard Model. It is vital to keep the trigger \gls{pt} thresholds as low as possible since the acceptance fraction of interesting processes, such as Higgs bosons decaying to tau leptons, decreases rapidly with increasing \gls{pt} thresholds.

The trigger is usually organized in at least two levels: a hardware trigger based on analogue electronics and logic in \glspl{FPGA} and \glspl{ASIC} followed by a software trigger running on a computer farm. The hardware trigger in the two LHC experiments reduces the event rate from the bunch-crossing rate of \SI{40}{\MHz} to the order of hundreds of \si{\kHz}. It only has a few microseconds to make a decision and must be located on or close to the detector. The software trigger uses sophisticated reconstruction algorithms and makes a trigger decision on the timescale of seconds. After the software trigger the event rate to offline storage is a few hundred \si{\Hz}.

In order to maintain low trigger thresholds, high efficiency and low rates, the two LHC experiments are developing methods to use tracking information in the hardware trigger. This has until now not been possible because of insufficient bandwidth to read out the tracker and resources to process the tracker data fast enough for the trigger decision. 

In this paper we compare two pattern recognition methods that can be used in the hardware trigger for fast hit filtering of tracking data. In section \ref{sec:tracktrig} we present the motivation and challenge of using tracking in the trigger. In section \ref{sec:simulation} we present the simulation framework. In sections \ref{sec:Hough} and \ref{sec:PatternMatching} we present the two methods for fast hit filtering: pattern matching using \gls{AM} chips and the Hough transform. Section \ref{sec:HitFilterComparison} presents the results from simulation and compares the two methods. A summary is given in section \ref{sec:summary}.

\section{Triggering with tracks}
\label{sec:tracktrig}
The tracker is the most granular detector system in the collider experiment. It consists of 10--20 layers, located at increasing distance from the interaction point, equipped with silicon detector modules. The layers at short distance from the interaction point are the most granular and are equipped with pixellated sensors giving two-dimensional information of where high energetic particles cross the sensor. Sensors with strips are used at large distance from the interaction point. These are less granular than the pixel sensors and give only one-dimensional information but requires less electronics to be read out. Two-dimensional information can be obtained from the strip layers using two layers of silicon strip sensors in close distance that are rotated by an angle.

To read out data from the tracker requires high bandwidth. In current trackers the services needed to meet the required bandwidth exceed the available space and power budget. Hence, methods to reduce the quantity of data to be read out are needed. This is done by introducing data reduction on-detector as developed by CMS or using readout in \glspl{RoI} seeded by the muon and calorimeter triggers, as developed by ATLAS. (In ATLAS, an \gls{RoI} is defined as a region in the track parameter phase-space that spans a maximum of \SI{10}{\%} of the tracker volume.)

The processing of tracker data is done in two main steps. The first is to identify hits from high-\gls{pt} particles originating from primary interactions. This is done with coarse resolution in order to save time and computing resources. The filtered hits with full resolution are forwarded to a  fitting step where the track parameters are determined. The performance of the hit filtering step is essential to keep the number of fits at a low level. The hit filtering step must have a high efficiency not to lose potentially interesting events. 

The two methods studied in this paper are the Hough transform and pattern matching with \gls{AM} which both can be implemented in hardware. The study is influenced by the development in ATLAS but the performed studies are general and valid for similar problems. We have decided to limit the study to an eight layer system to make it easier to compare the two methods.   
An eight layer system is a reasonable choice for a hardware based system since limited bandwidth constraints and computing resources still do not allow all potential layers to be used.

\section{Simulation}
\label{sec:simulation}
\begin{figure}[t]
\centering 
\includegraphics[width=.45\textwidth]{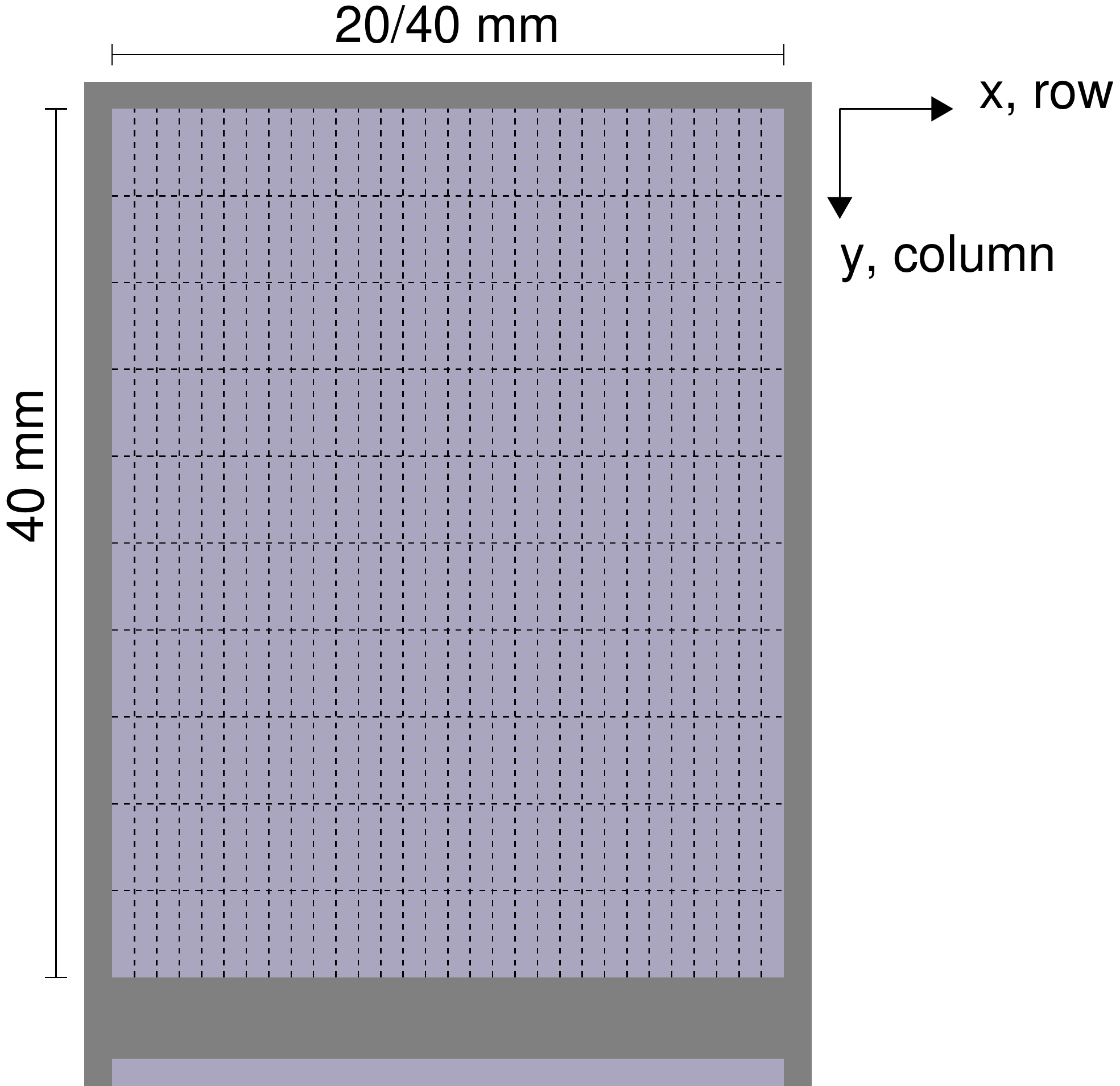}
\includegraphics[width=.45\textwidth]{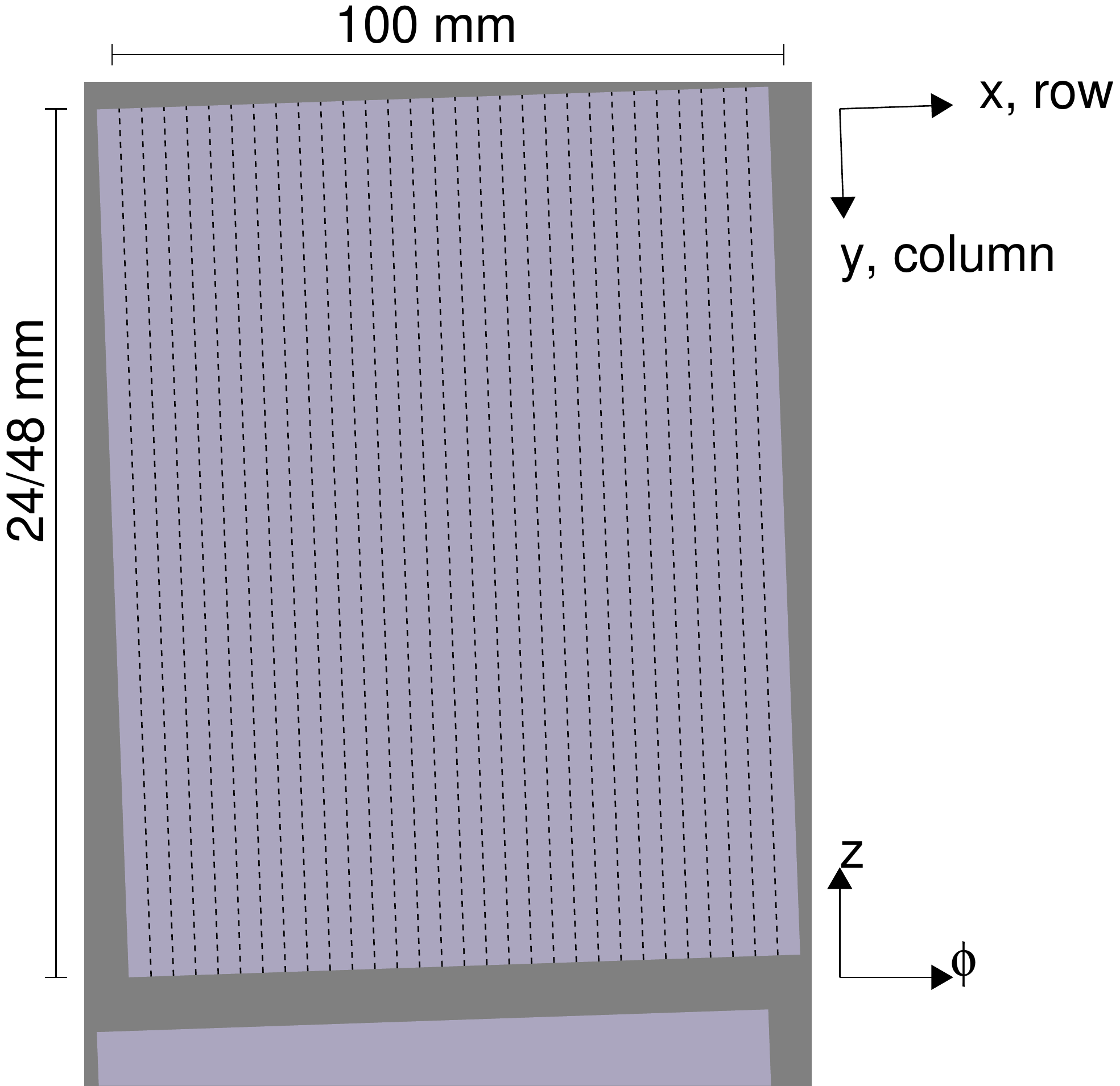}
\caption{\label{fig:sensors} Schematic depiction of the pixel (left) and strip (right) sensors. The dashed lines show the orientation of the individual strips and pixels (not to scale).}
\end{figure}

\begin{figure}[t]
\centering 
\includegraphics[width=.6\textwidth]{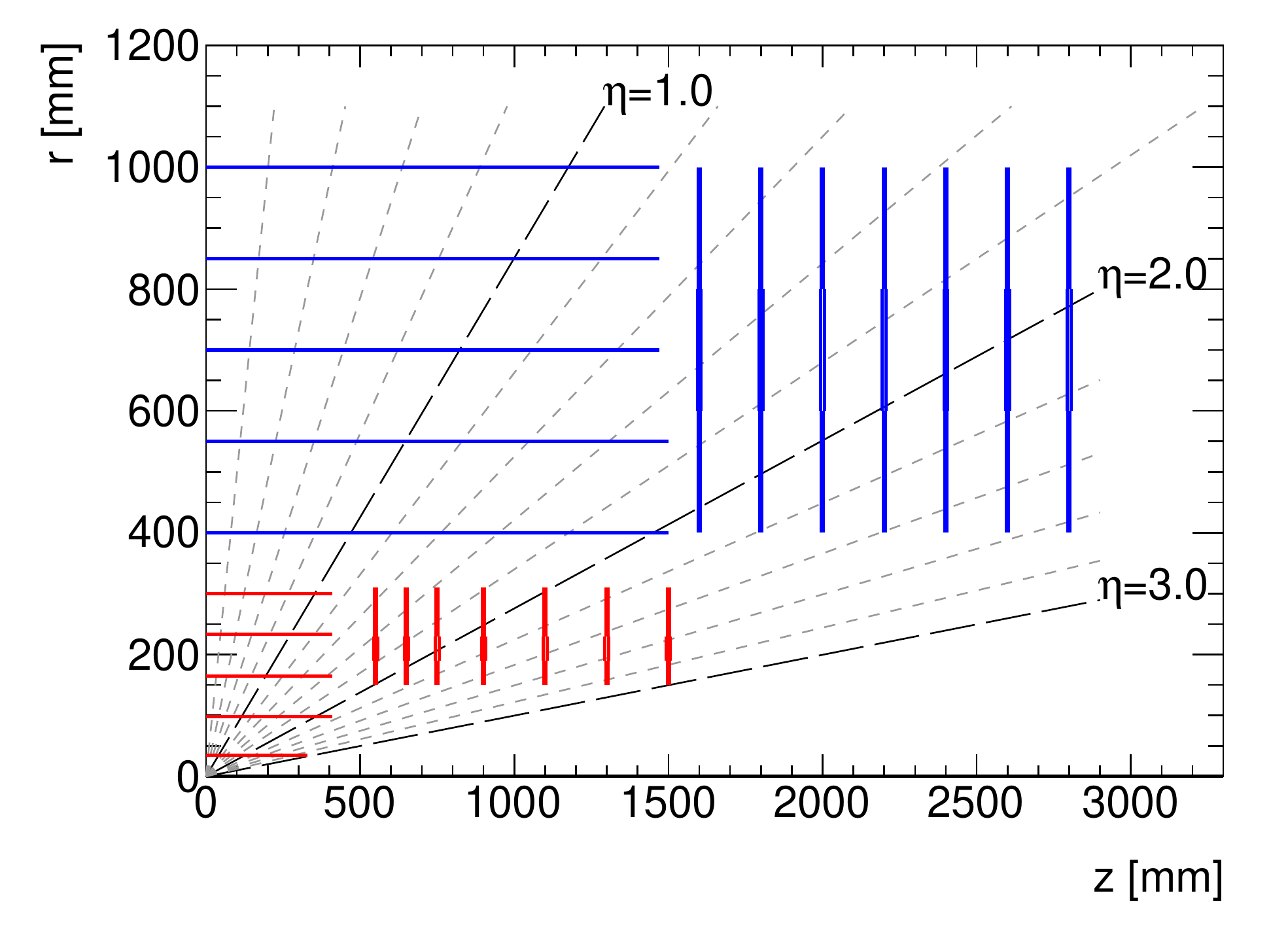}
\caption{\label{fig:DetectorLayout} Layout of the detector with pixel sensors in red and strip sensors in blue.}
\end{figure}

\begin{figure}[t]
\centering 
\includegraphics[width=.48\textwidth]{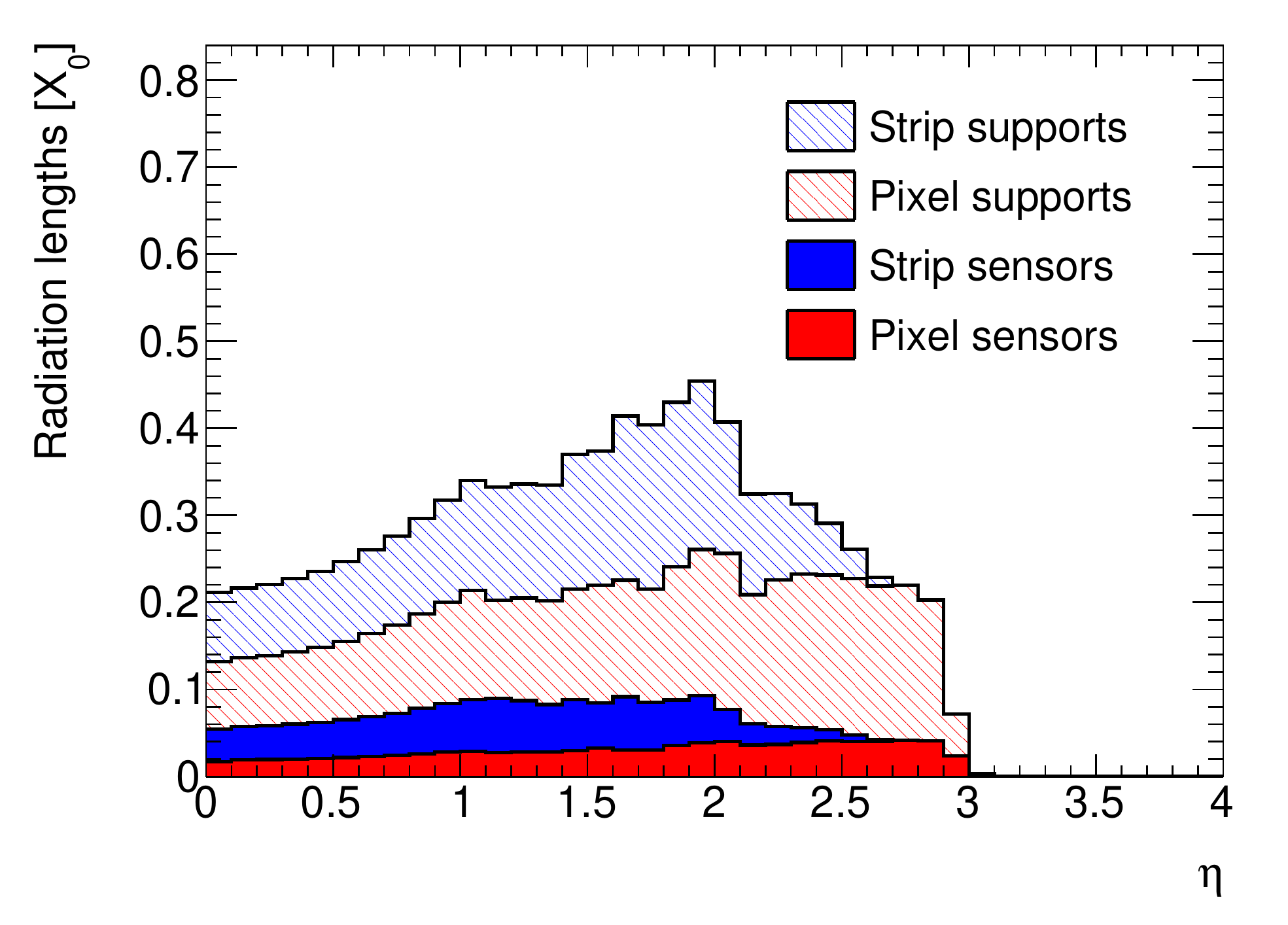}
\includegraphics[width=.48\textwidth]{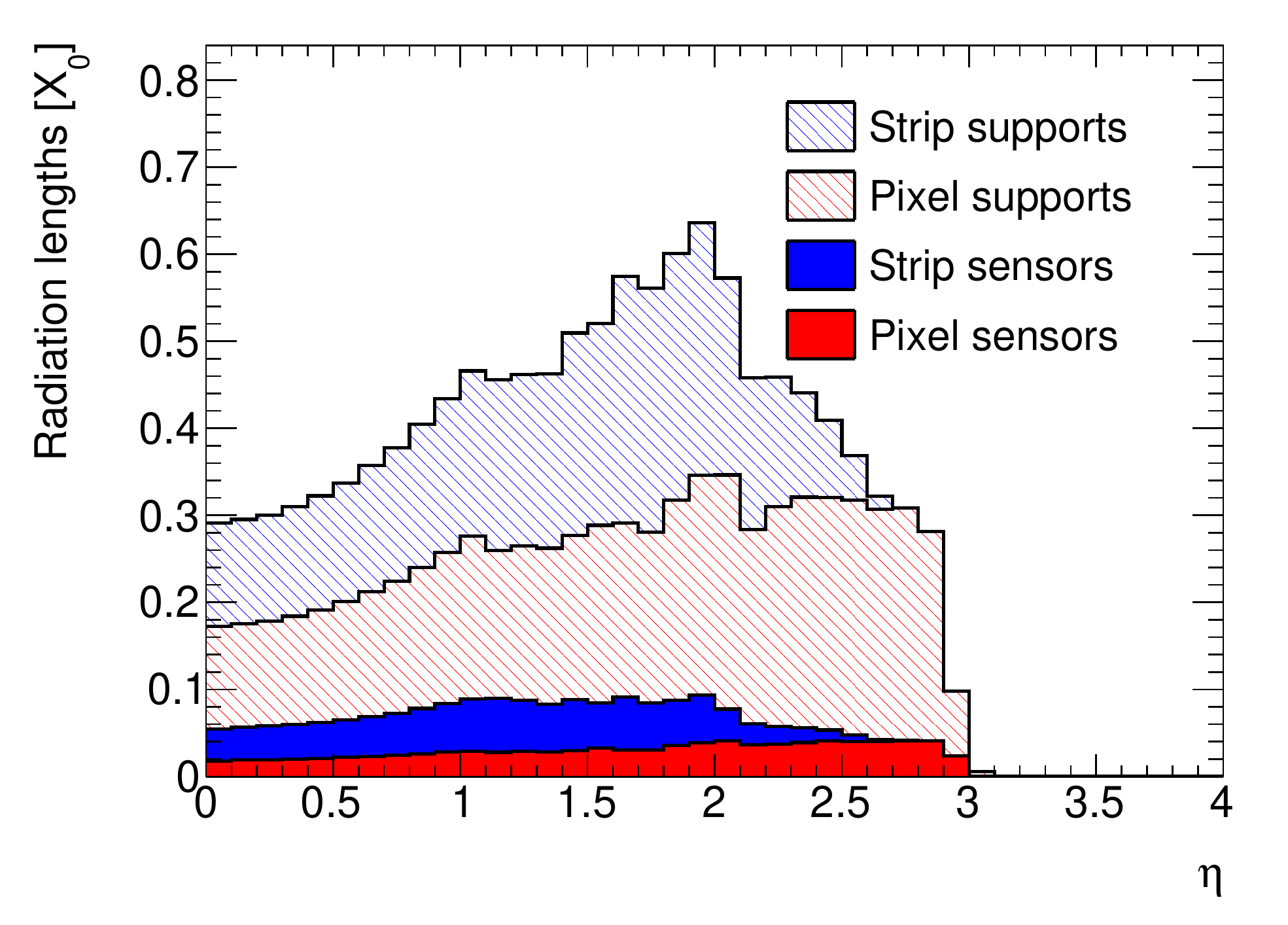}
\caption{\label{fig:RadiationLength} Number of radiation lengths traversed by geantino particles as a function of pseudorapidity in the nominal setup (left) and with more dense support material (right).}
\end{figure}

A generic tracking detector, similar in layout to those of the ATLAS and CMS experiments, is modeled in \geant~\cite{Geant4}. 
The detector has a barrel section, with sensors arranged in cylindrical shells around the beam axis, and endcaps with disks of sensors in the transverse plane.
A schematic view of the layout is shown in figure~\ref{fig:DetectorLayout}, and scans of the radiation length is shown in figure~\ref{fig:RadiationLength}.
The sensors are modeled as rectangular boxes of silicon, \SI{320}{\um} thick in the radial (beam axis) direction in the barrel (endcaps).
The hit positions in the sensors are recorded modulo the readout segmentation, i.e. the local $(x,y)$ coordinates are turned into discrete hit positions of $(\text{\textit{row}}, \text{\textit{column}})$.
Figure~\ref{fig:sensors} shows the layout and local coordinates of the sensors on the support material.
The sensors with one dimensional segmentation, so-called \emph{strip} detectors, divide the local $x$ coordinate in steps of \SI{80}{\um}.
Sensors with two-dimensional segmentation, so-called \emph{pixel} detectors, divide the local $x$-coordinate by \SI{25}{\um} and the local $y$-coordinate by \SI{250}{\um}.
The barrel region contains five layers with pixel sensors and five double layers of strip sensors, with details given in table~\ref{tab:BarrelGeometry}. 
In addition to the sensors, all layers contain material in the form of \SI{3}{\mm} thick carbon supports.
The double layers have strip sensors on each side of the stave rotated with a stereo angle of \SI{40}{\milli\radian}. The stereo angle provides fine resolution for offline data, but at the hardware track trigger stage this information is not available.

Single muons were generated directly in \geant~using a particle gun, while minimum bias events were generated with \pythia~\cite{Pythia8} using the \texttt{SoftQCD:inelastic} setting for proton beams with a \SI{14}{\TeV} center of mass energy. 
The propagation of the particles through the detector was simulated in \geant~with the \texttt{FTFP\_BERT} physics list.
The muons were generated with impact parameters sampled from uniform distributions of $|z_0|<\SI{150}{\mm}$ and $|d_0|<\SI{2}{\mm}$. 
The transverse momentum ranged from 4 to \SI{400}{\GeV} and was sampled from a uniform distribution of $1/p_{\text{T}}$.
In addition, the muon $\eta$ and $\phi$ was restricted to one RoI, $0.1\leq\eta_0\leq0.3$ and $0.3\leq\phi_0\leq0.5$.

The \geant~simulation is followed by a \emph{digitization} step. In this step a single muon event is overlaid with 0 to 260 minimum bias events to emulate various degrees of pile-up. Muons with primary hits outside the RoI are discarded. The hits are merged by adding together the total energy deposited in each pixel or strip and requiring that the total energy deposited corresponds to more than \SI{1}{\femto\coulomb} of collected electric charge. It is these merged hits that are referred to as \emph{hits} in the rest of the paper. Hits coming directly from \geant -generated muons are labeled as signal and everything else as background. 

The number of connected strips or pixels in $\phi$ that have a hit, the so-called \emph{cluster width}, is also calculated for each hit. The cluster width is later used in the Hough transform and the AM pattern matching to ignore hits with cluster width larger than 3. The physical motivation for this is that the high-\gls{pt} tracks are straighter and produce smaller clusters than the more bent low-\gls{pt} track from minimum bias.

\section{Hough transform}
\label{sec:Hough}

The Hough transform was originally developed for particle physics to detect particle tracks in photographic plates from bubble chambers in the late 1950s~\cite{hough-bubble}. Since then the method has been generalized and is commonly used to detect features that can be represented by a few parameters, such as lines and circles, in a wide range of imaging applications~\cite{hough_duda}.
The Hough transform is actively used in offline tracking, e.g. in ALICE~\cite{hough-alice}, and is gaining interest with recent advancements in parallel computing; see for example \cite{hough-halyo}. To be used in triggering, the Hough transform must be implemented in hardware, for instance using \glspl{FPGA}. In this paper, the Hough transform is simulated using \textsc{C++}, but a very similar implementation has been written in \textsc{OpenCL} and run on an \gls{FPGA}.

The Hough transform calculates all combinations of parameters consistent with each data point and casts votes in a histogram-like object called an \emph{accumulator}, that has one dimension for each parameter in the feature representation. The coordinates of the points in the accumulator with the most votes corresponds to candidate features. 

The track of a charged particle in the transverse plane of a uniform magnetic field, e.g. in a tracking detector, is described by a circular arc. If the interaction vertex is constrained to the origin of the coordinate system and the \gls{RoI} is small enough in $\phi_0$ (such that $\sin(x) \approx x$), the track through a point with polar coordinates $(r,\varphi)$ can be described by

\begin{equation}
  \label{eq:hough}
  A\frac{qB}{p_\text{T}} = \frac{\phi_0 - \varphi}{r},
\end{equation}

\noindent where $A\approx\SI{1.5e-4}{\GeV\per\lightspeed\per\mm\per\tesla}$ is the unit conversion constant if $r$ is measured in millimeters, $B$ is the magnetic field in \si{\tesla}, $p_\text{T}$ is the transverse momentum of the track in \si{\GeV\per\lightspeed}, and $q$ is the sign of the electric charge.

The accumulator is implemented as a two-dimensional histogram with $A\frac{qB}{p_\text{T}}$ on one axis and $\phi_0$ on the other. The accumulator is filled by entering the hit coordinates 
into equation~(\ref{eq:hough}) and sweeping $\phi_0$ over the range defined by the \gls{RoI} to get the $A\frac{qB}{p_\text{T}}$ coordinate. However, each bin does not simply contain the number of hits. Instead, the bin consists of an 8-bit number to keep track of which of the eight layers have been hit in the small parameter space that it spans. It also has a list of all the hits in the bin. After the accumulator has been filled, bins with less than 6 layers hit are discarded. The surviving bins are track candidates with rough track parameters $\phi_0$ and \gls{pt} given by the location of the bin in the accumulator and a list of hits that can be sent to the track fitter.

The proton-proton collisions in ATLAS at the \gls{HL-LHC} are expected to be spread out uniformly over \SI{300}{\mm} along the beam line, i.e. $z$-axis, making the \gls{RoI} very wide in $z_0$~\cite{itktdr}. Since the Hough transform, as implemented here, is looking at a projection in the transverse plane, this makes it difficult to separate tracks with similar $\phi_0$ and \gls{pt} that originate from different points along the beam line. The two conventional approaches of separating tracks in $z_0$ using the Hough transform is either to parameterize the track in $z_0$ as well as $\phi_0$, or make a second pass of the Hough transform to find lines in the $r$-$z$ plane after finding track candidates in the transverse plane. The problem with the first approach is that it requires to sweep both $\phi_0$ and $z_0$, which adds another dimension to the accumulator and increases the number of computations needed quadratically. The problem with the second approach is that the Hough transform in the $r$-$z$ plane would have to be performed after completing the Hough transform in the transverse plane and would increase the execution time proportionally to the number of hits passing the first stage, approximately \SI{30}{\%} for single muons embedded in 200 minimum bias events, since the hits are read out serially. Also, the track finding efficiency is found to be poor due to the low resolution of the strip layers in $z$.

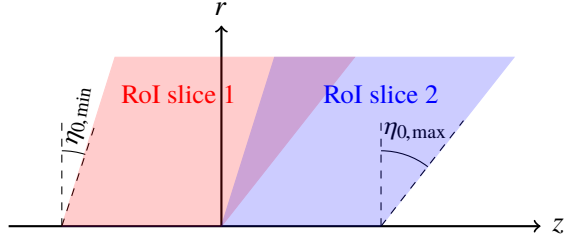
\begin{figure}[ht]
\centering
  \begin{tikzpicture}[scale=0.7]
    \draw[->,thick] (-4,0) -- (6,0) node[right] {$z$}; 
    \draw[->,thick] (0,0) -- (0,3.8) node[above] {$r$};
    \draw[black] (-3,0) pic{carc=72:90:1};
    \draw[black,dashed] (-3,0) -- (-3,2.0); 
    \draw[black,dashed] (-3,0) -- (-2.35,2.0); 
    \draw[color=black](-2.66,2.1) node[rotate=80] {\small$\eta_{0,\text{min}}$}; 
    \draw[black] (3,0) pic{carc=52:90:1};
    \draw[black,dashed] (3,0) -- (3,2.0); 
    \draw[black,dashed] (3,0) -- (4.55,2.0); 
    \draw[color=black](3.65,1.8) node[rotate=0] {\small$\eta_{0,\text{max}}$};
    \draw[fill,red,opacity=0.2] (-3,0) -- (-2,3.2) -- (2.5,3.2) -- (0,0) -- (-3,0);
    \draw[red](-0.8,2.5) node {\small RoI slice 1};
    \draw[fill,blue,opacity=0.2] (0,0) -- (1,3.2) -- (5.5,3.2) -- (3,0) -- (-3,0);
    \draw[blue](3,2.5) node {\small RoI slice 2};
  \end{tikzpicture}
\caption{The RoI can be sliced up in $z_0$ to reduce the occupancy in the Hough accumulator. In this case the RoI is split into two slices.}
\label{fig:houghsplit}
\end{figure}

Our approach to help separating tracks in $z_0$ is to slice up the \gls{RoI} in several parts in $z_0$ as illustrated in figure~\ref{fig:houghsplit} and fill separate accumulators for each slice by sorting the hits according to their $z$-coordinate. The \gls{RoI} slices overlap with each other slightly in $z$ because of the $\eta_0$ range and due to the non-zero detector resolution in $z$, especially for the strips. The $z_0$-slicing approach is found to reduce the number of hits passing the Hough transform by \SI{70}{\%} without affecting the track finding efficiency for single muons embedded in 200 minimum bias events.

The number of bins in $A\frac{qB}{p_\text{T}}$ and $\phi_0$, as well as the number of slices in $z_0$, need to be tuned to find a configuration that provides a high-enough track finding efficiency while suppressing as much of the low-\gls{pt} background as possible. The optimum number of bins in the accumulator depends strongly on the detector geometry, the detector resolution, and the layer configuration used. The optimal number of $z_0$-slices is mostly affected by the pile-up conditions.

The Hough transform can be implemented in hardware using an \gls{FPGA}. The \gls{FPGA} receives the hits serially and fills the whole $\phi_0$ range and all $z_0$-slices in parallel. There are $N_z$ accumulators, one for each $z_0$-slice, with $N_{p_\text{T}}$ bins in $A\frac{qB}{p_\text{T}}$ and $N_{\phi}$ in $\phi_0$ consisting of \SI{8}{\bit}, one for each layer. The total memory requirement for the accumulators is $N_z\times N_{p_\text{T}}\times N_{\phi} \times \SI{8}{\bit}$. When all hits have been processed, bins with at least 6 hits in unique layers are considered track candidates with rough track parameters $\phi_0$, \gls{pt}, and $z_0$. At this point full-resolution hits need to be associated with the track candidates and sent to the track fitter. This is expected to be done in a separate FPGA.

The available hardware constrains the size of the accumulator used. For the particular geometry and layer configuration used in this study (described in section \ref{sec:simulation}), the accumulator typically has $N_z = 6$, $N_{p_\text{T}} = 120$, and $N_{\phi} = 300$, amounting to a memory requirement of \SI{216}{\kilo\byte}. The low latency requirement of a hardware track trigger prevents the use of external memory; hence, the accumulator must be stored internally in the \gls{FPGA}.

The vertex constraint imposed when deriving equation (\ref{eq:hough}) is necessary to reduce the computational complexity and the time needed to perform the Hough transform. Without the vertex constraint each hit has to be paired with every other hit and the amount of computation grows as the square of the number of hits in the event. However, imposing a vertex constraint has an important drawback: it limits the ability to find tracks with large impact parameters, i.e. those which originate far from the beam line, especially if using detector layers located close to the interaction point.

\section{Pattern matching}
\label{sec:PatternMatching}

Pattern matching is a commonly employed technique in particle track finding; a review of its use can be found in~\cite{Mankel}. 
At the LHC, pattern matching is used by the ATLAS and CMS experiments both in track triggers and offline reconstruction to reduce the computational load on
track fitting algorithms.
A pattern usually consists of a set of hit positions in the detector, which correspond to the track of a high energy particle. 
The actual hits in the detector can be compared to these patterns, and only the hits that match a pattern are propagated to the track fitting step. 
The pattern matching step can be made very fast by using a subset of the available detector hit positions, a coarser resolution, or both.
ATLAS is currently commissioning a hardware track trigger, the Fast TracKer (FTK)~\cite{FTK}, which performs the pattern matching in \gls{AM} chips~\cite{AMchips}.
This technology has been used in a wide variety of applications outside of particle physics, such as network switches and artificial neural networks.
The AM chips can compare a detector hit to the stored patterns in parallel, making for a very fast system. 
This method is considered for a fast hardware track trigger in the HL-LHC era by both ATLAS and CMS. A constraint on the design of the trigger comes from the number of patterns
that can be stored on a single AM chip, which also limits the number of layers that can be used. The current estimation for the AM chips under development for the ATLAS
upgrade is that they can hold about half a million patterns using eight layers. Here we assume that two such chips can be used to cover a phase space of $0.2\times0.2$ in $\Delta\eta_0\times\Delta\phi_0$ based on the cost and power requirements of these chips.

In this study, the patterns consist of a set of hit positions with coarse resolution in a subset of eight silicon detector layers. 
The coarse resolution hit positions, so-called \emph{superstrips}, are formed by groups of contiguous silicon strip or pixel elements.
A unique superstrip index (SSID) is created from the module and superstrip number, and stored in the pattern.
The patterns are generated by simulating single muons and recording the hit positions in the selected layers. 
About 30 million muons are used to build a bank of a few million patterns. 
A large sample is needed to get good efficiency, but some events produce the same pattern which is why the bank typically contains less than five million patterns. 
The patterns are ordered by their \emph{use count}, i.e. how many events produce that same pattern. 
To respect the hardware constraints, only the first one million patterns are used to match hits in a separate sample of muons overlaid with minimum bias events.
During the matching all hits in the detector are converted to their superstrip index and compared to each entry in the pattern bank.
Each pattern can have multiple detector hits within the superstrips which are propagated to the track fit. 

The patterns use eight separate layers and a match requires seven of these layers to have a superstrip hit. 
Since a realistic detector can have inefficiencies due to the lack of coverage, e.g. dead space between sensors, the efficiency of the pattern matching is increased by using patterns with \emph{wild cards}. 
If a muon lacks a hit in any of the layers during the pattern generation, the layer can be marked as a wild card for this pattern. 
In the pattern matching, a pattern with a wild card always considers that layer to have been hit.
With a maximum of two wild cards per pattern it is possible to have a matching pattern with only five real hits in separate layers, which increases the rate of fake matches from combinations of background hits
and degrades the track fit performance. For these reasons, patterns with two wild cards are required to have all eight layers hit to be matched, i.e. six real hits.

The number of patterns in a bank is mainly driven by the size of the superstrips, as the number of possible patterns grow with smaller sizes. 
Using larger superstrips is not an ideal option, since it increases the amount of fake matches. A compromise can be found by using \emph{don't care} (DC) bits.
A DC bit on the SSID in the pattern bank effectively doubles the size of the superstrip, as it disregards the value of the least significant bit in the SSID.
Many patterns share the same SSID in several layers and/or have neighboring SSIDs. These patterns can then be combined into one pattern with one or more DC bits.
Since a DC bit can be set individually for each pattern and layer, this creates a bank of patterns with variable resolution.
This reduces the total number of patterns in the bank but only marginally increases the number of fake matches.

\section{Comparison of the hit filtering}
\label{sec:HitFilterComparison}
The track fitter which follows after the hit filtering will have to perform one fit for each hit combination. Therefore, the number of combinations of hits after the pattern recognition stage is an important measure of performance. Both the Hough transform and the AM approach group hits together. In the Hough transform the hits are grouped into bins in the accumulator, while in the AM pattern matching, each matched pattern is associated with a group of hits. The total number of fits $N_\text{fits}$ needed is calculated by taking the product of the number of hits in each layer and then summing over the number of groups:

\begin{equation}
  \label{eq:160706}
  N_\text{fits} = \sum\limits_g \left( \prod\limits_{l=1}^{8} n_{g,l} \right) \text{,}
\end{equation}

\noindent where $n_{g,l}$ is the number of hits in layer $l$ of group $g$.

The other important measure of performance is the track finding efficiency. A muon track is considered to be successfully found if at least 6 out of 8 hits from the primary muon are found in unique layers.  This definition is motivated by that the subsequent track fit, e.g. the one planned for the ATLAS Phase-II hardware track trigger, is foreseen to fit tracks with 8, 7, and 6 out of 8 hits.  The efficiency is defined as the number of events with a muon found divided by the total number of events. One thing to note is that the track finding efficiency of the AM pattern matching is limited by the number of patterns that can fit in the hardware. The Hough transform, on the other hand, can be configured to provide almost any efficiency within the same hardware. However, a higher efficiency comes with the cost of a larger number of possible hit combinations. For the results presented here, the Hough transform was tuned to provide similar efficiency as the AM method.

In section \ref{sec:results_nominal} we present and compare the performance of the hit filtering for AM pattern matching and the Hough transform in terms of the number of hit combinations after the hit selection, the efficiency of finding muon tracks, and how many hits originating from the true muon that pass the hit filters. In section \ref{sec:results_more_material} we present the effect of adding \SI{50}{\%} more material, and in section \ref{sec:results_module_dropout} we present the effect of randomly turning off \SI{5}{\%} or \SI{10}{\%} of the modules.

\subsection{Study of nominal performance}
\label{sec:results_nominal}

\begin{figure}[t]
\centering 
\includegraphics[width=.8\textwidth]{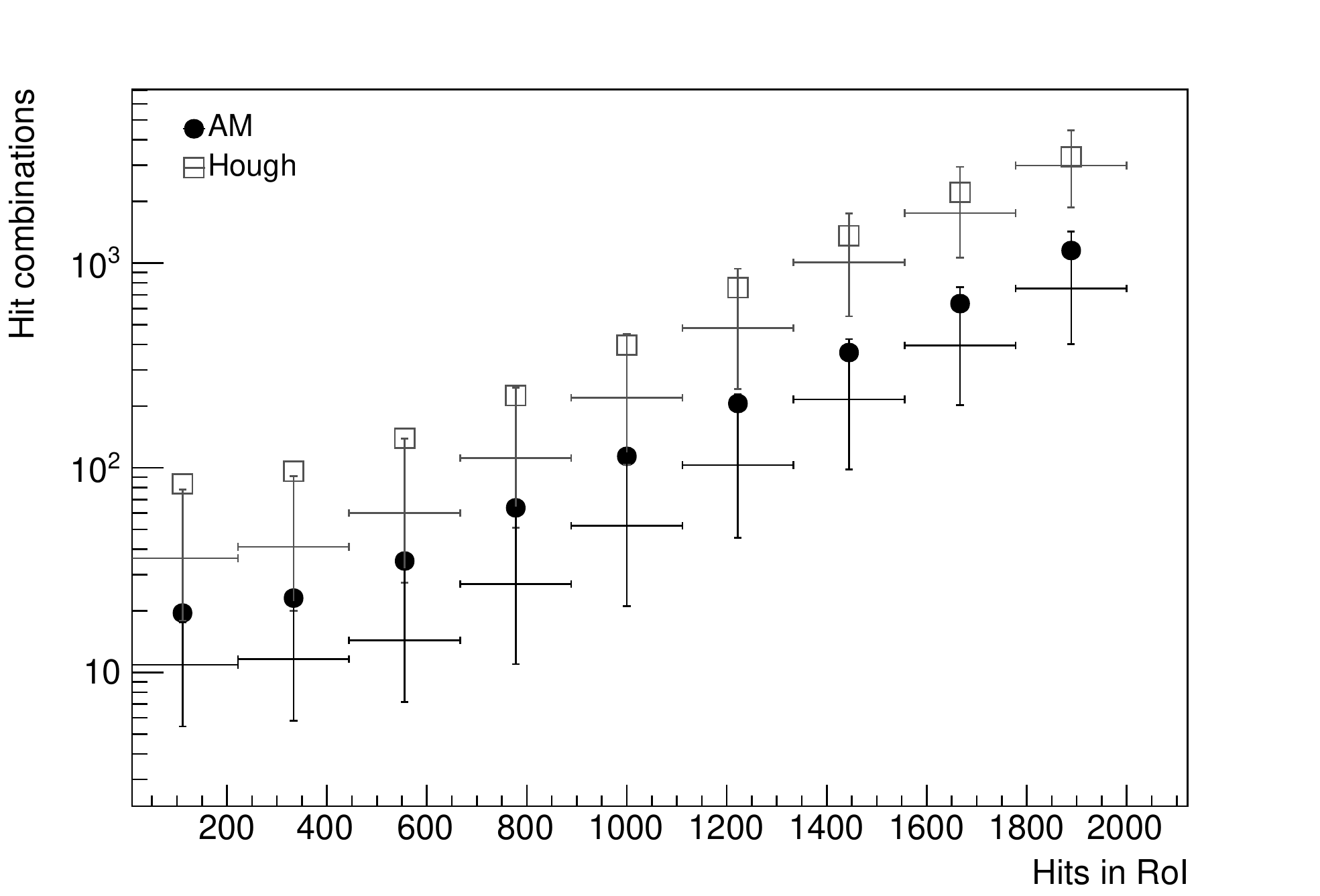}
\caption{\label{fig:combinations} Number of hit combinations after the AM pattern matching and Hough transform as a function of the number of input hits. The circles and squares mark the average number of combinations for the AM method and the Hough transform respectively. The crosses mark the \SI{50}{\%} (median) with the vertical bars marking the \SI{25}{\%} and \SI{75}{\%} percentiles of the distribution.}
\end{figure}

\begin{figure}[t]
\centering
\includegraphics[width=.8\textwidth]{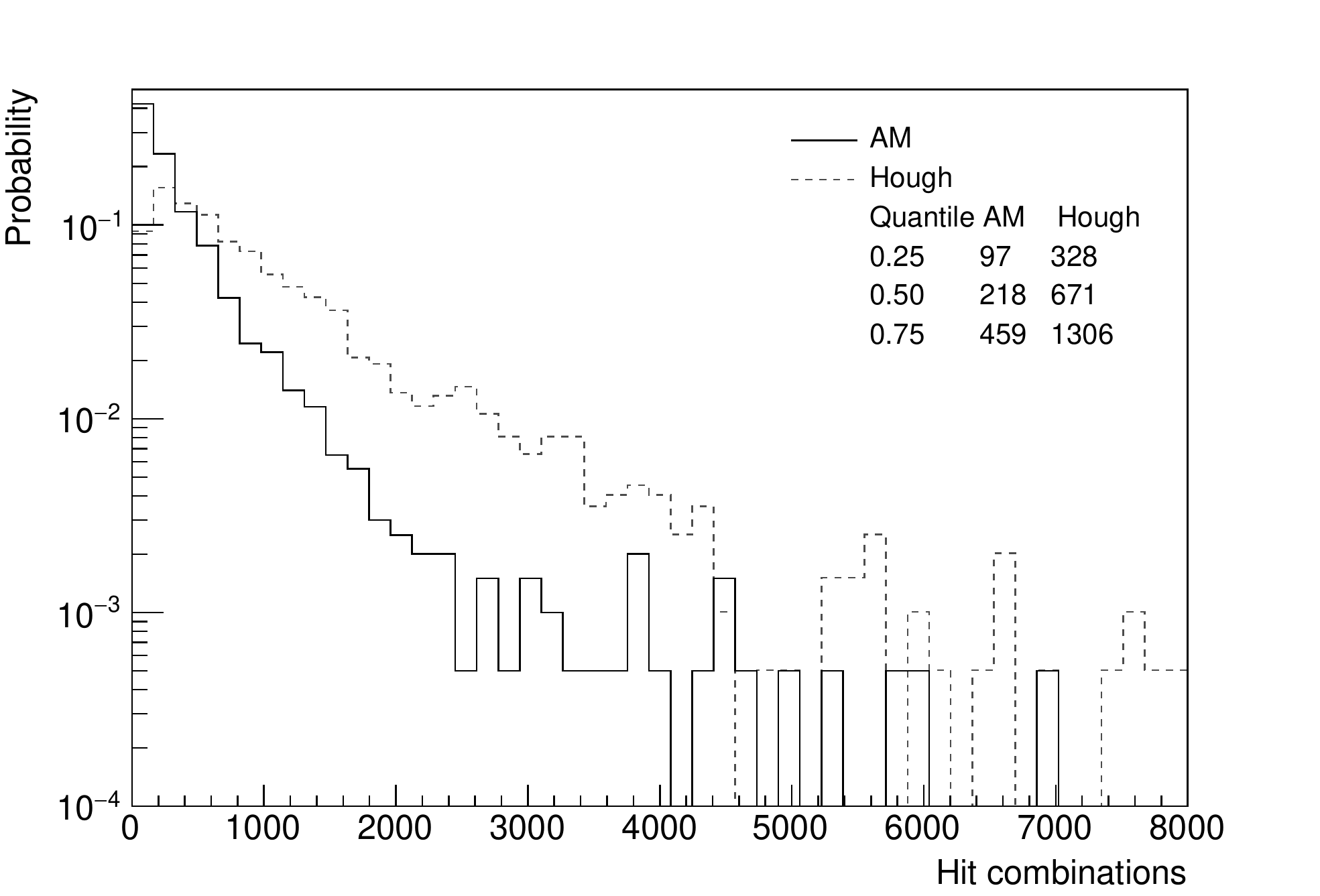}
\caption{\label{fig:combinations_pu200} Distribution of the number of hit combinations after the AM pattern matching and Hough transform for single muons embedded in 200 minimum bias events.}
\end{figure}

Figure \ref{fig:combinations} shows the number of hit combinations as a function of the number of hits in the \gls{RoI} when both methods are providing roughly the same track finding efficiency. The number of hit combinations required for the Hough transform is higher than that required for the AM method. Both show the same exponential increase as a function of the number of hits. The distribution of the number of hit combinations are not symmetrical, and have tails that extend to a very high number of combinations, as seen in figure~\ref{fig:combinations_pu200}.

\begin{figure}[t] 
\centering 
\includegraphics[width=.8\textwidth]{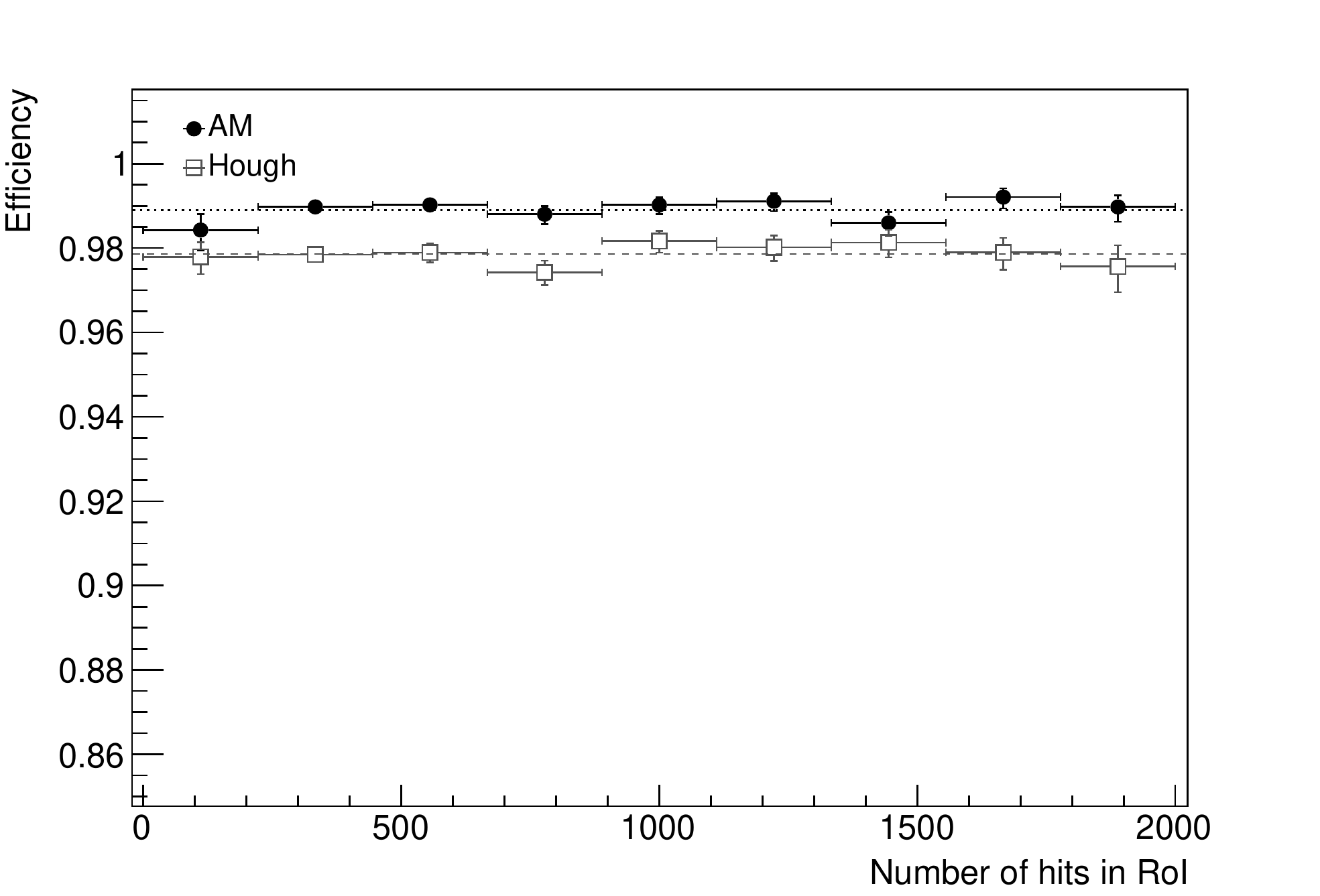}
\caption{\label{fig:efficiency_nominal}  Muon track finding efficiency of the AM pattern matching and Hough transform as a function of the number of hits in the RoI. The dotted and dashed lines show the average efficiency for the AM method and Hough transform respectively. }
\end{figure}

\begin{figure}[t]
\centering 
\includegraphics[width=.8\textwidth]{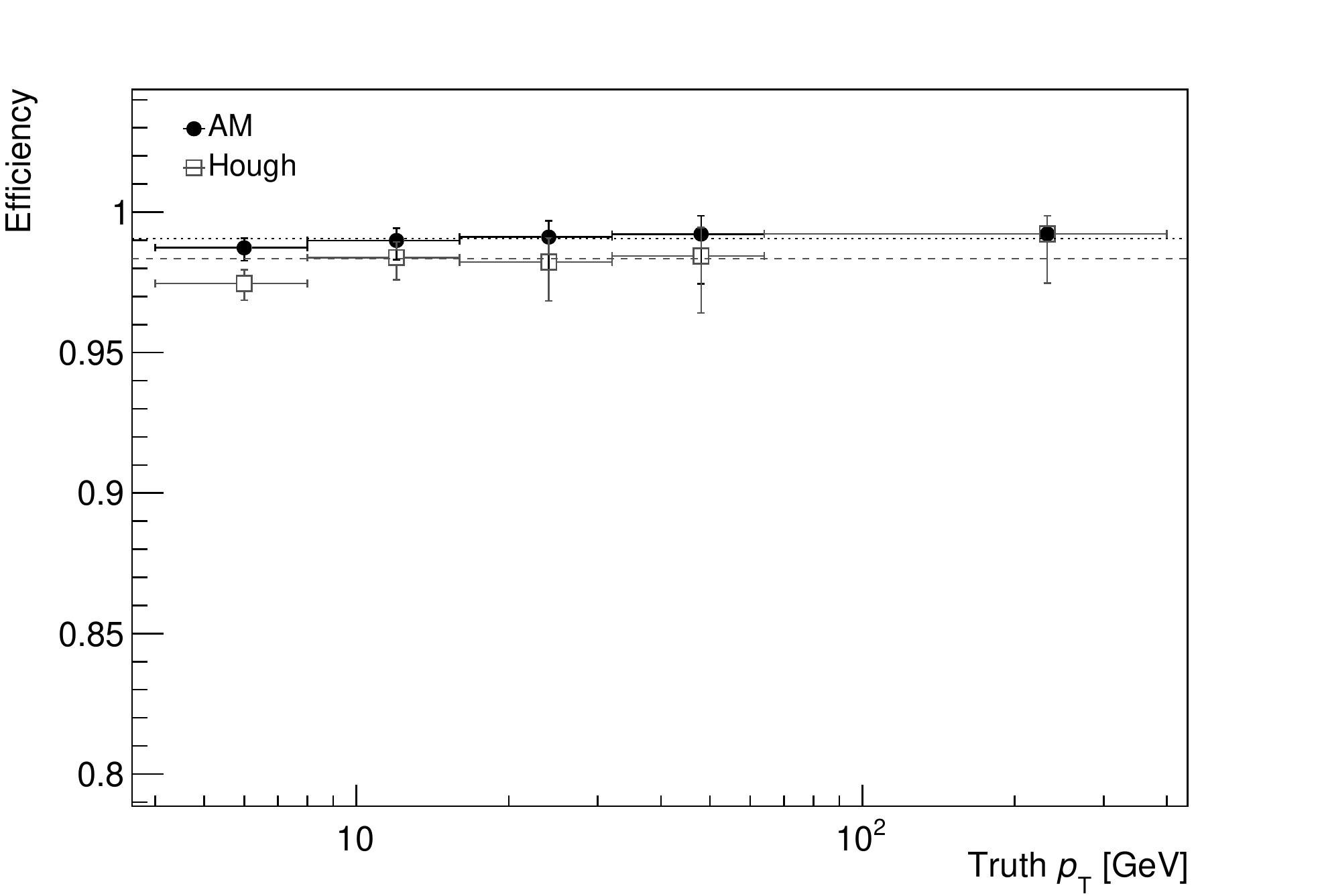}
\caption{\label{fig:turnon_nominal} Muon track finding efficiency of the AM pattern matching and the Hough transform as a function of the true muon \gls{pt} for single muons embedded in 200 minimum bias events. The dotted and dashed lines show the average efficiency for the AM method and Hough transform respectively. The efficiency for the AM pattern matching and the Hough transform happen to overlap in the highest \gls{pt} bin.}
\end{figure}

The efficiency of finding tracks from primary muons is plotted against the number of hits in figure~\ref{fig:efficiency_nominal} and against the true muon $p_\text{T}$ in figure~\ref{fig:turnon_nominal}. Both methods have a flat efficiency as a function of the number of hits in the \gls{RoI}. Having high efficiency for leptons with \gls{pt} in the 4--\SI{20}{\GeV} range is important to maintain a low trigger \gls{pt} threshold and performing track-based isolation. As shown in figure \ref{fig:turnon_nominal}, the difference between the track finding efficiency at 4--\SI{8}{\GeV} and 64--\SI{400}{\GeV} is only a fraction of a percent for the \gls{AM} method. The Hough transform shows a larger difference, approximately \SI{1.5}{\%}.

\begin{figure}[t]
\centering 
\includegraphics[width=.48\textwidth]{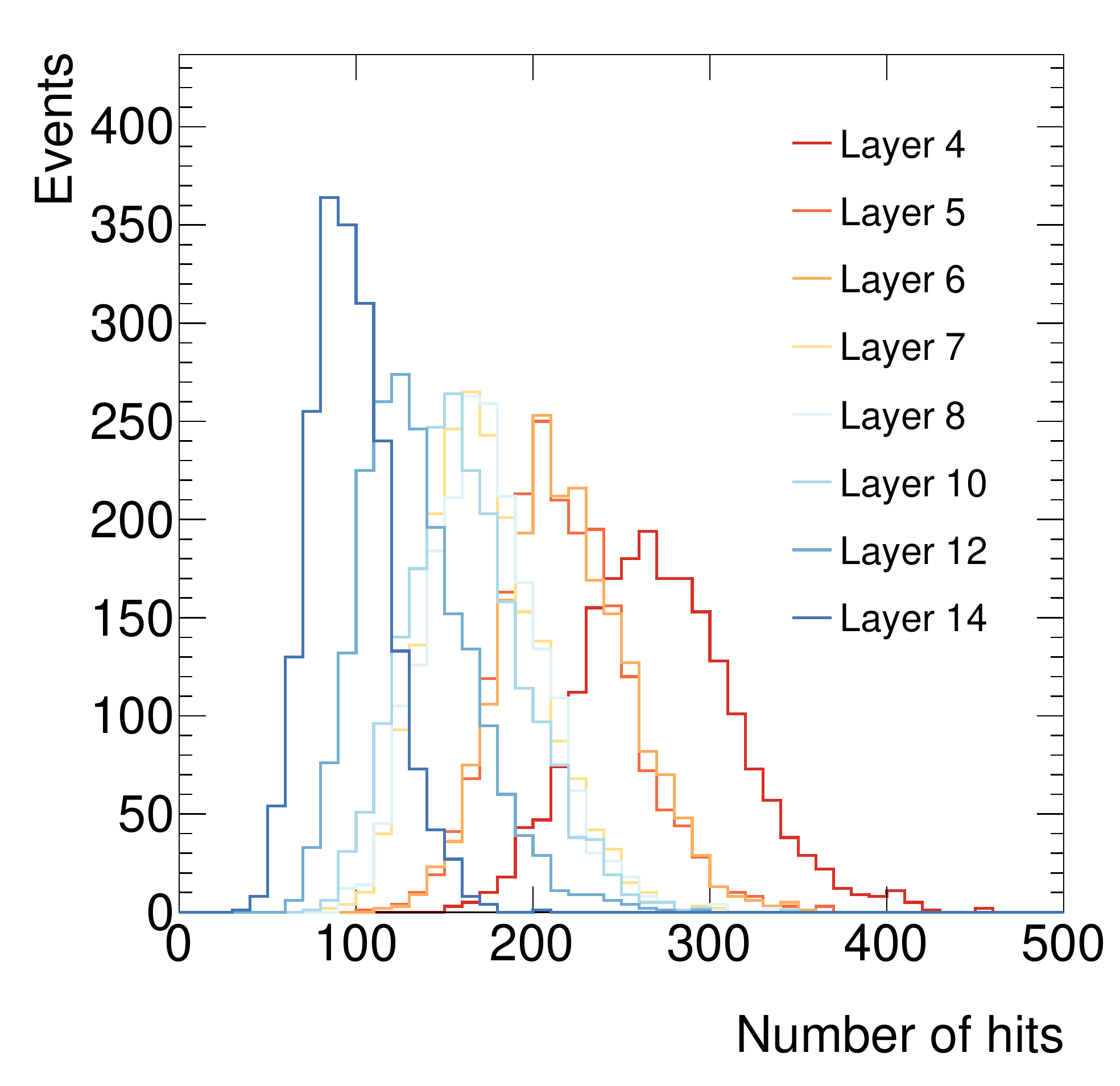}
\caption{\label{fig:hitsInRoI} Number of hits in the RoI for single muon events overlaid with 200 minimum bias events, before applying the AM pattern matching or Hough transform.}
\end{figure}

\begin{figure}[t]
\centering 
\includegraphics[width=.48\textwidth]{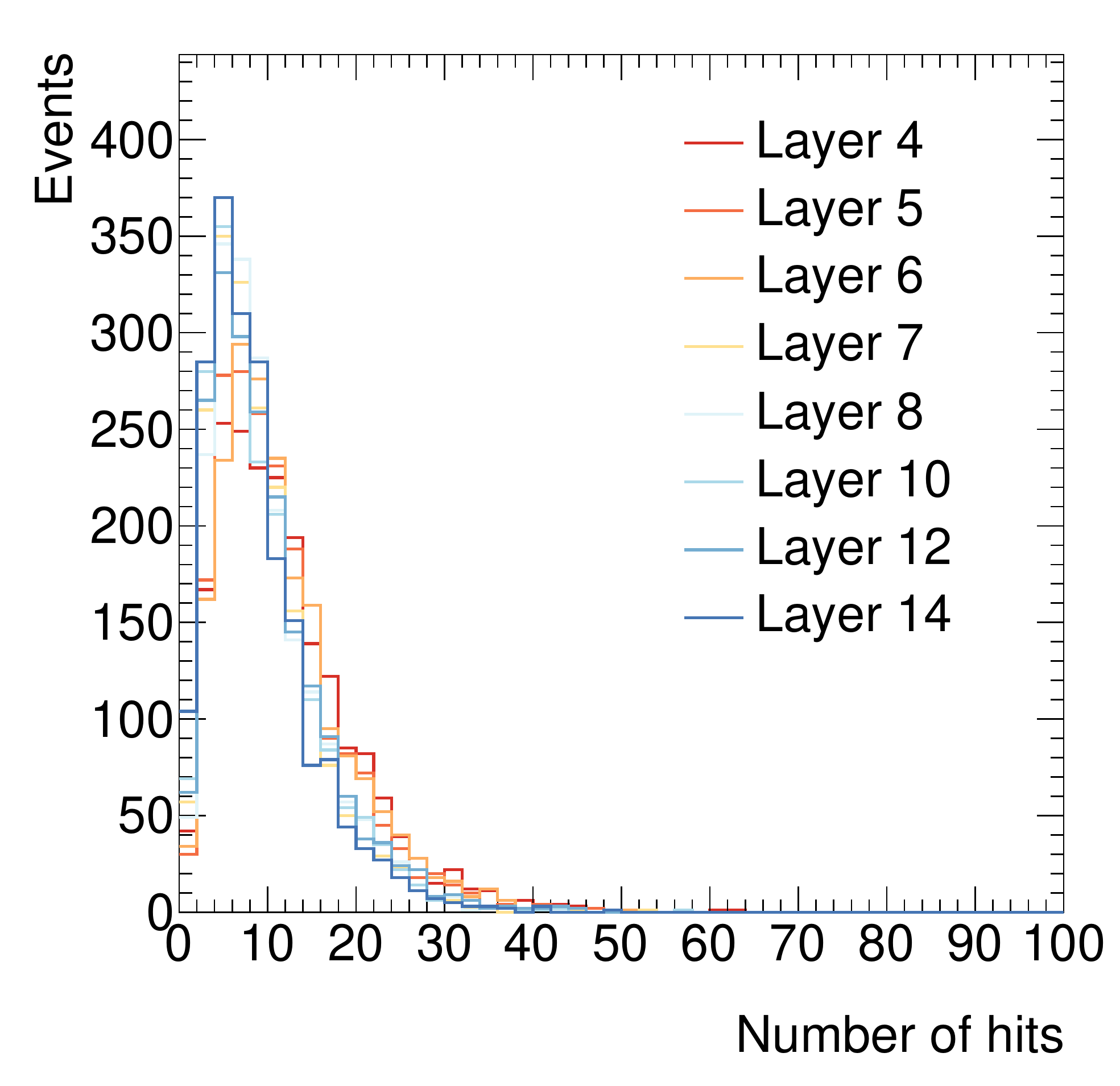}
\includegraphics[width=.48\textwidth]{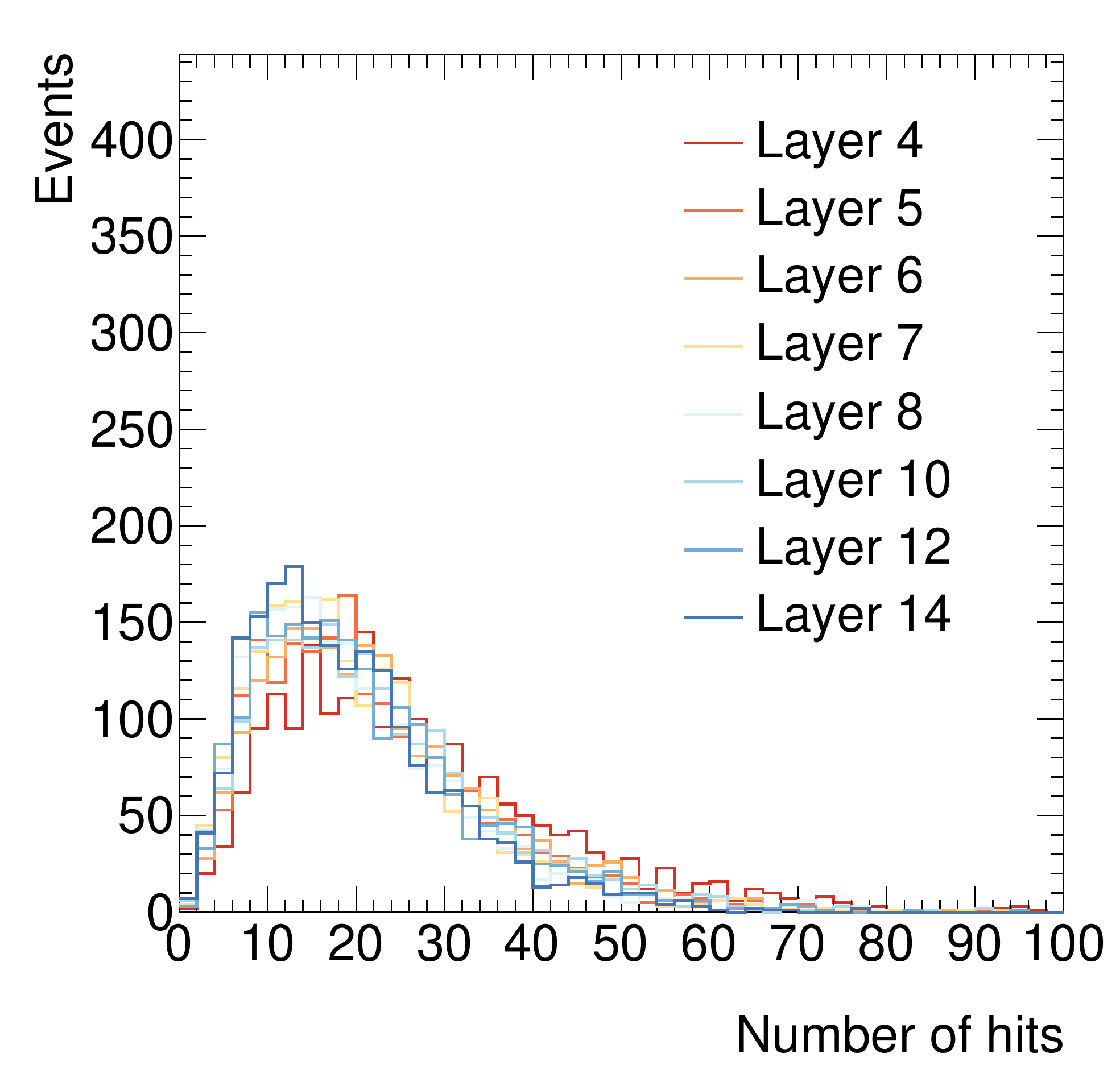}
\caption{\label{fig:hitsAfterFiltering} Number of hits in the RoI for single muon events overlaid with 200 minimum bias events, after applying the AM pattern matching (left) or Hough transform (right).}
\end{figure}

It is important to verify that the hit filtering methods are not biased to any particular layer. The inner layers are expected to have more hits since tracks from low-\gls{pt} charged particles are bent off by the magnetic field. As expected, the number of hits in each layer decrease with larger radius, as shown in figure~\ref{fig:hitsInRoI}. After the \gls{AM} pattern matching or Hough transform all layers have roughly the same number of hits, as seen in figure~\ref{fig:hitsAfterFiltering}. This is expected for tracks of high-\gls{pt} particles and verifies that neither method is biased towards any particular layer.

\begin{figure}[t]
\centering 
\includegraphics[width=.48\textwidth]{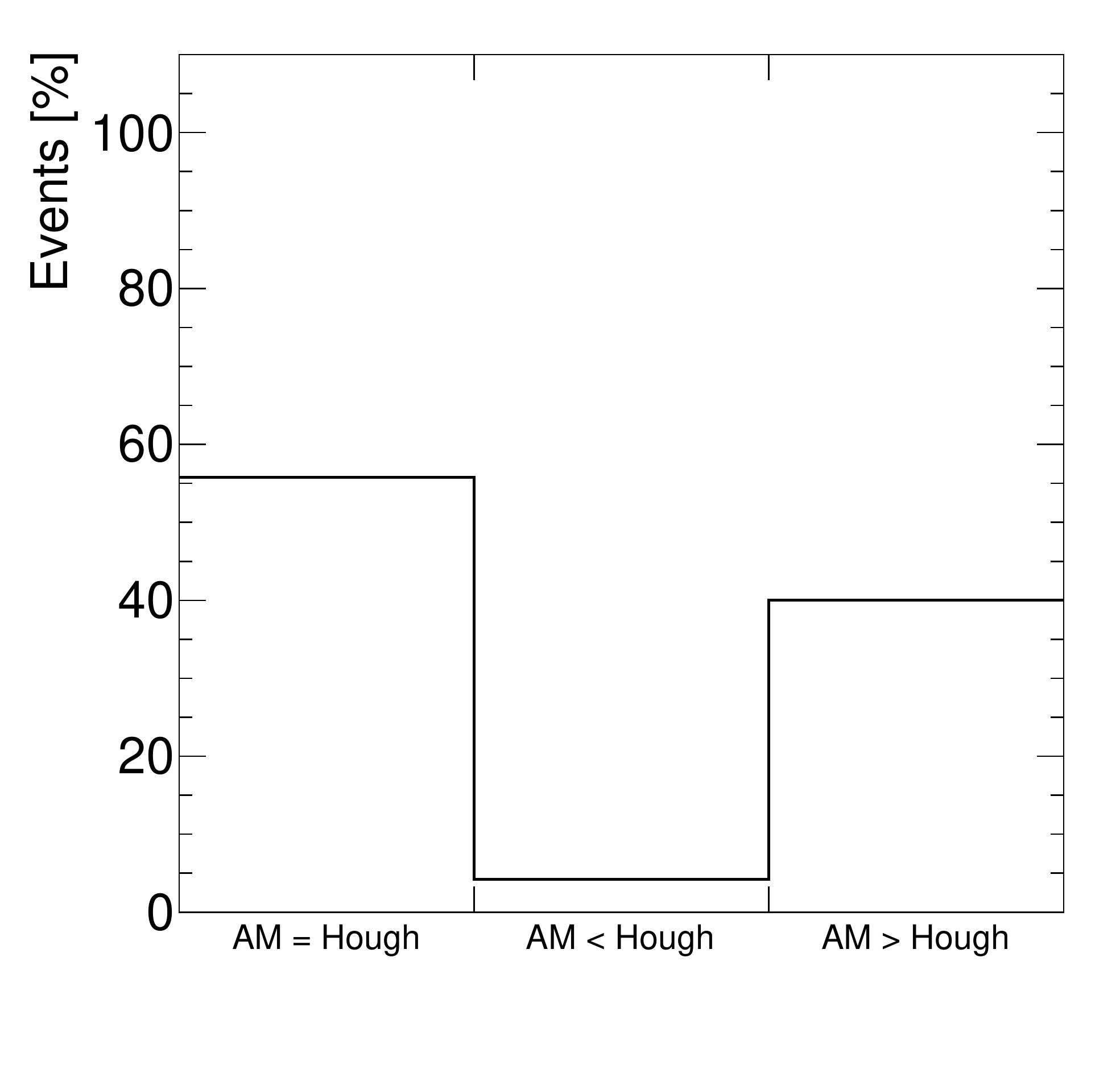}
\caption{\label{fig:muonHitsDifference} Fraction of events (in \si{\%}) where the AM pattern matching and the Hough transform differ in the number of primary muon hits found in the best pattern or group for single muon events overlaid with 200 minimum bias events.}
\end{figure}

\begin{figure}[t]
\centering 
\includegraphics[width=.48\textwidth]{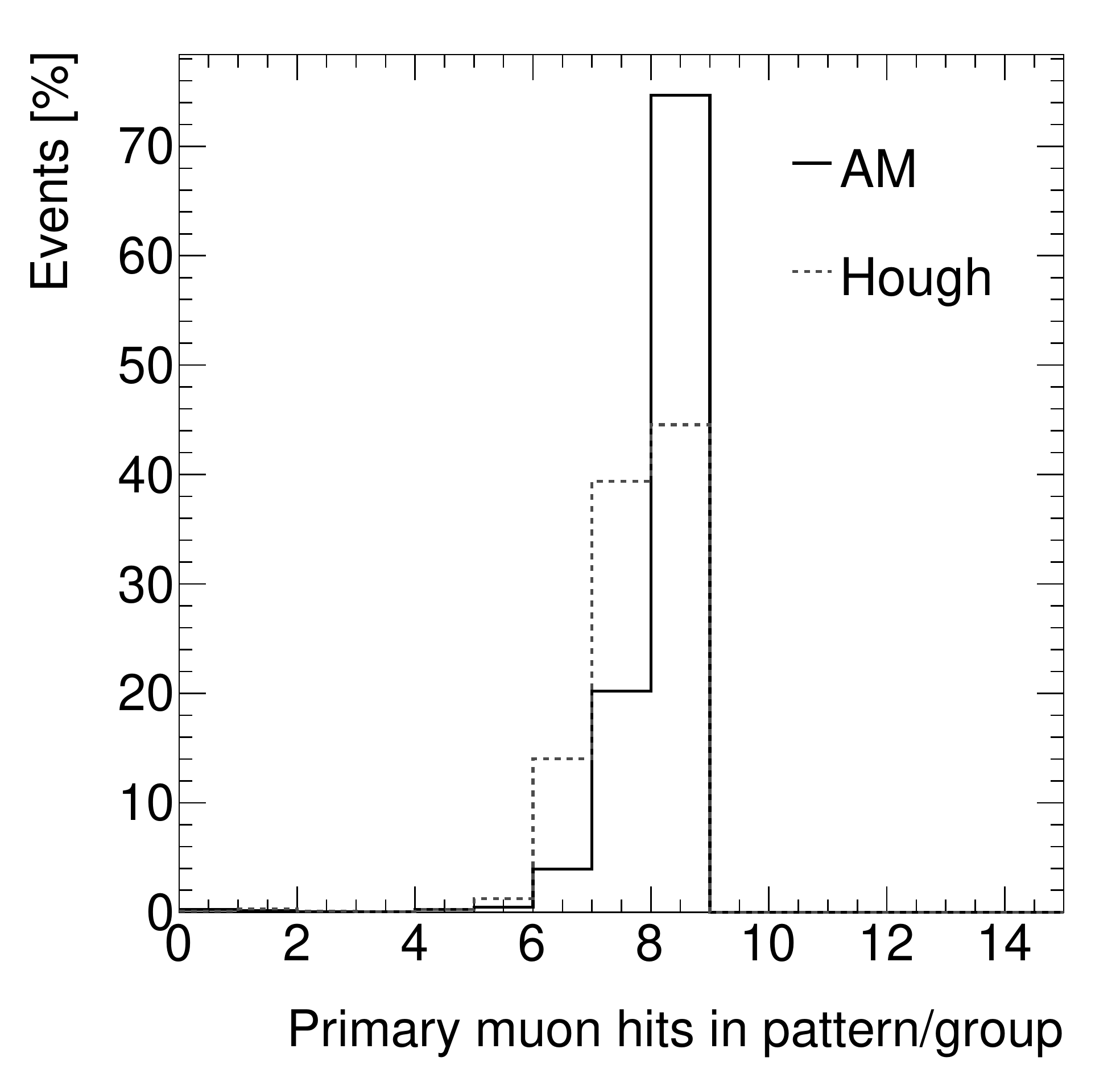}
\caption{\label{fig:nHitsCorrectlyAssigned} Number of primary muon hits found by the AM pattern matching and the Hough transform for single muon events overlaid with 200 minimum bias events.}
\end{figure}

The number of hits from primary muons sent to the fitter will have an effect on the resolution of the track parameters, with the best performance obtained when 8 hits can be used. Figure~\ref{fig:muonHitsDifference} shows the fraction of events where the AM and Hough transform find the same or different number of muon hits, and figure~\ref{fig:nHitsCorrectlyAssigned} shows the distributions of the number of muon hits found per event. The Hough transform and the AM pattern matching find the same number of muon hits in \SI{40}{\%} of the cases. In \SI{55}{\percent} of the events, the AM matching finds more muon hits than the Hough transform. The Hough transform finds more hits in only \SI{5}{\percent} of the events. This result clearly favors the AM pattern matching for the subsequent track fitting stage.

\subsection{The effect of detector material}
\label{sec:results_more_material}

\begin{figure}[t] 
\centering 
\includegraphics[width=.8\textwidth]{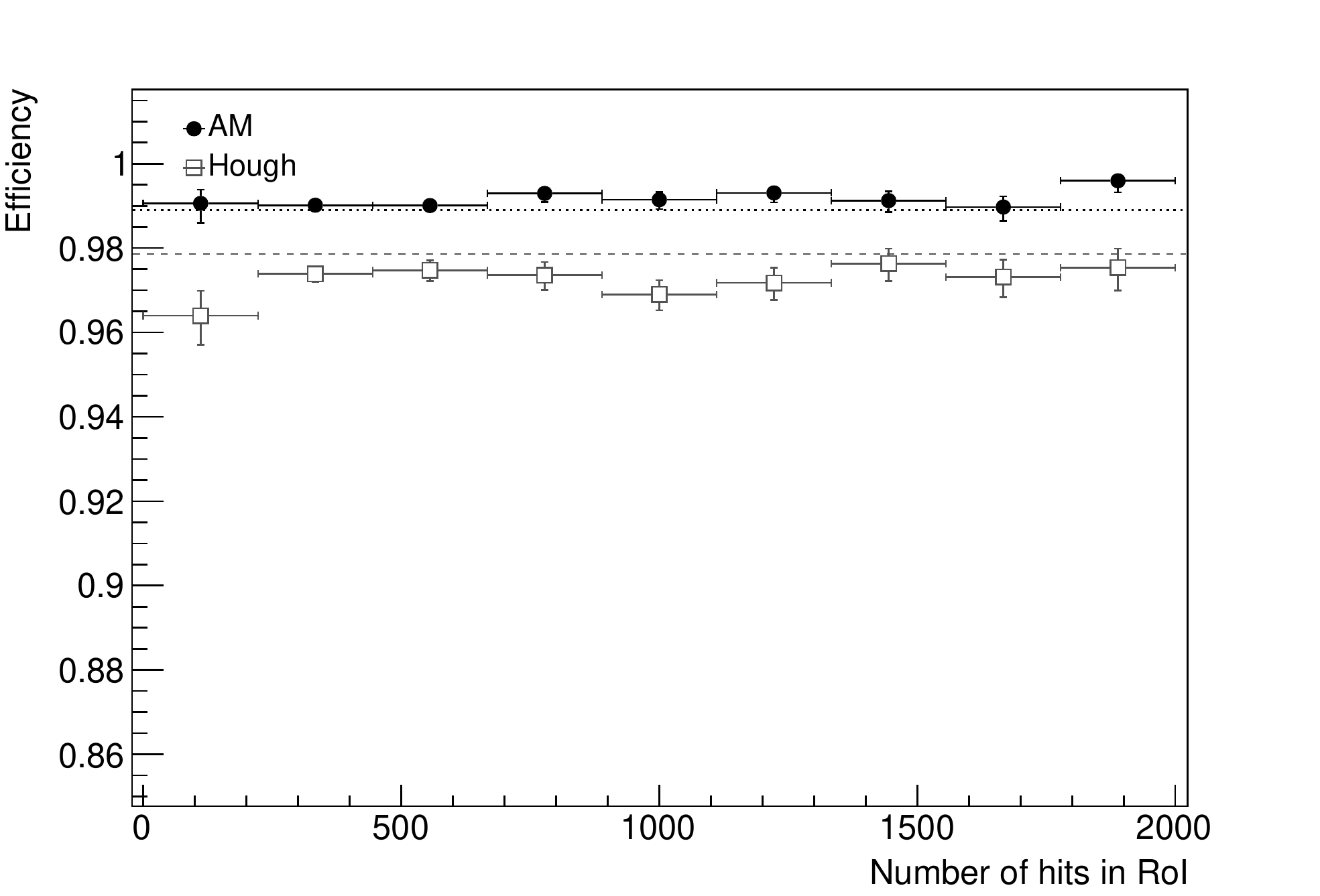}
\caption{\label{fig:efficiency_more_material}  Muon track finding efficiency of the AM pattern matching and Hough transform as a function of the number of hits in the RoI when adding \SI{50}{\%} more material. For reference, the dotted and dashed lines show the average nominal efficiency for the AM method and Hough transform respectively.}
\end{figure}

The effect on the track finding efficiency of adding \SI{50}{\%} more service material as a function of the number of hits in the \gls{RoI} is shown in figure \ref{fig:efficiency_more_material}. The AM methods shows a small overall increase in efficiency while the Hough transform shows an overall decrease.

\begin{figure}[t]
\centering 
\includegraphics[width=.8\textwidth]{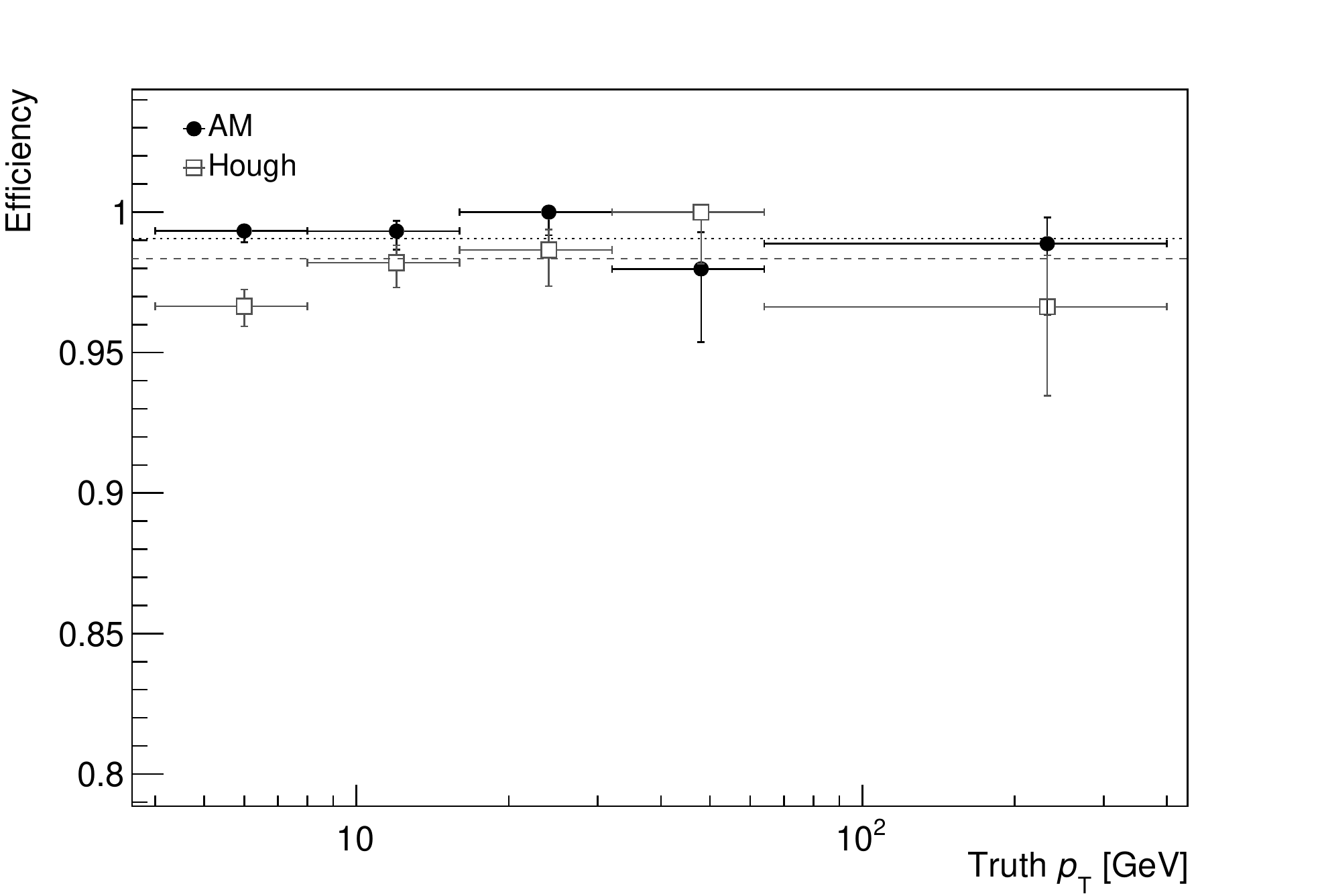}
\caption{\label{fig:turnon_more_material} Muon track finding efficiency of the AM pattern matching and the Hough transform as a function of the true muon \gls{pt} for single muons embedded in 200 minimum bias events when adding \SI{50}{\%} more material. For reference, the dotted and dashed lines show the average nominal efficiency for the AM method and Hough transform respectively.}
\end{figure}

Looking at the \gls{pt} dependence, shown in figure \ref{fig:turnon_more_material}, the \gls{AM} method has a small increase in efficiency, although the effect is compatible with statistical fluctuation. The Hough transform shows a decrease in efficiency of approximately \SI{1}{\%} at low and high \gls{pt}, although these effects are also within statistical uncertainties. Note also that the Hough transform should be able to re-gain efficiency by re-tuning the accumulator configuration at the cost of more hits surviving the selection.

\subsection{The effect of dropping detector modules}
\label{sec:results_module_dropout}

\begin{figure}[t] 
\centering 
\includegraphics[width=.8\textwidth]{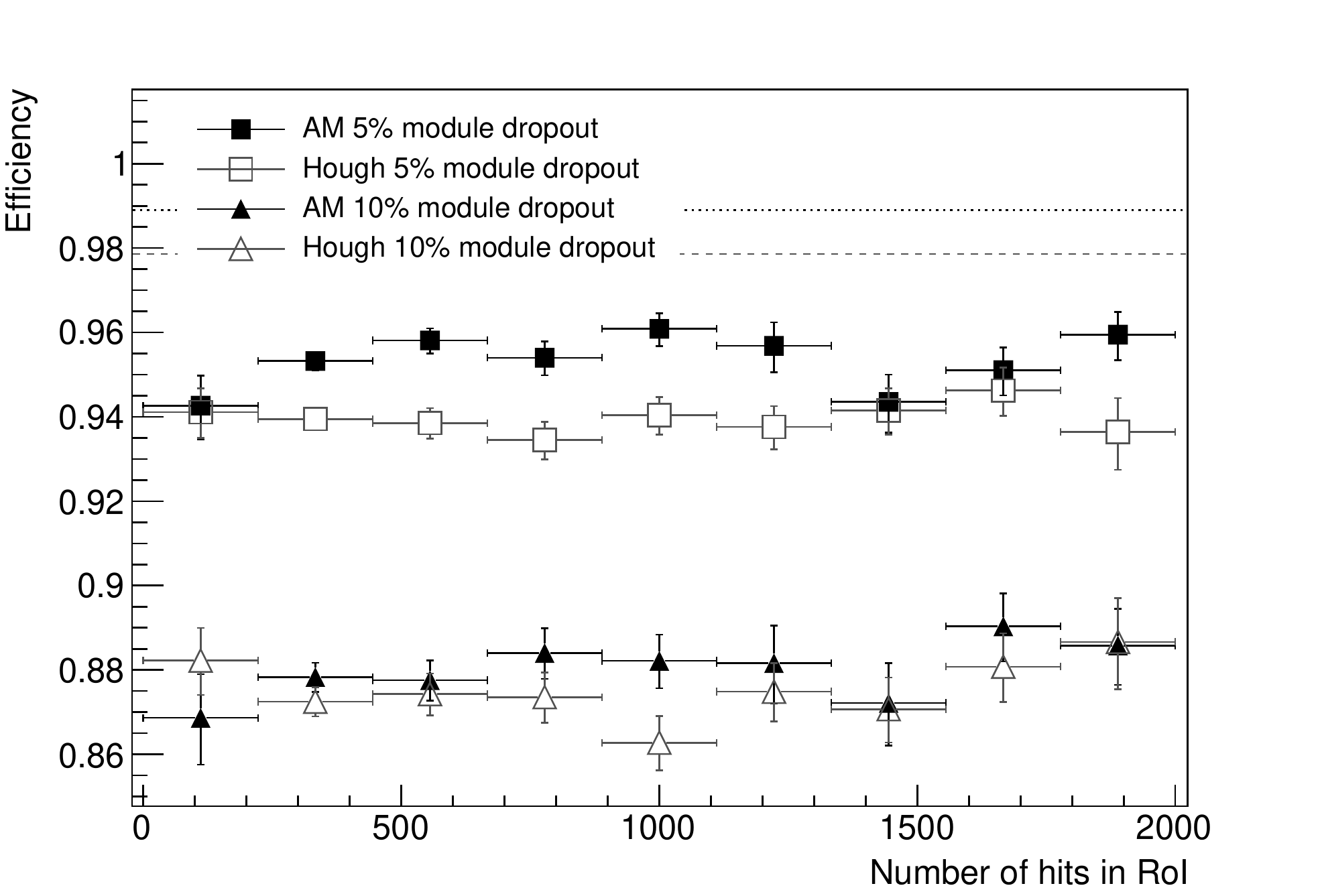}
\caption{\label{fig:efficiency_module_dropout} Muon track finding efficiency of the AM pattern matching and Hough transform as a function of the number of hits in the RoI when randomly dropping \SI{5}{\%} and \SI{10}{\%} of the detector modules. For reference, the dotted and dashed lines show the average nominal efficiency for the AM method and Hough transform respectively.}
\end{figure}

Based on experience with the current trackers at the LHC, a failing module is a more likely cause for loss of efficiency than a general reduction in quantum efficiency of the silicon sensors. In this study, inefficiencies are introduced by randomly removing single detector modules from the simulation. Figure \ref{fig:efficiency_module_dropout} shows the muon track finding efficiency as a function of the number of hits per RoI. Removing \SI{5}{\%} of the modules reduces the efficiency to approximately \SI{95}{\%} for both the Hough transform and the AM pattern matching, with a slightly lower efficiency for the Hough transform. Removing \SI{10}{\%} of the modules reduces the efficiency to approximately \SI{87}{\%} for both methods.

\begin{figure}[t]
\centering 
\includegraphics[width=.8\textwidth]{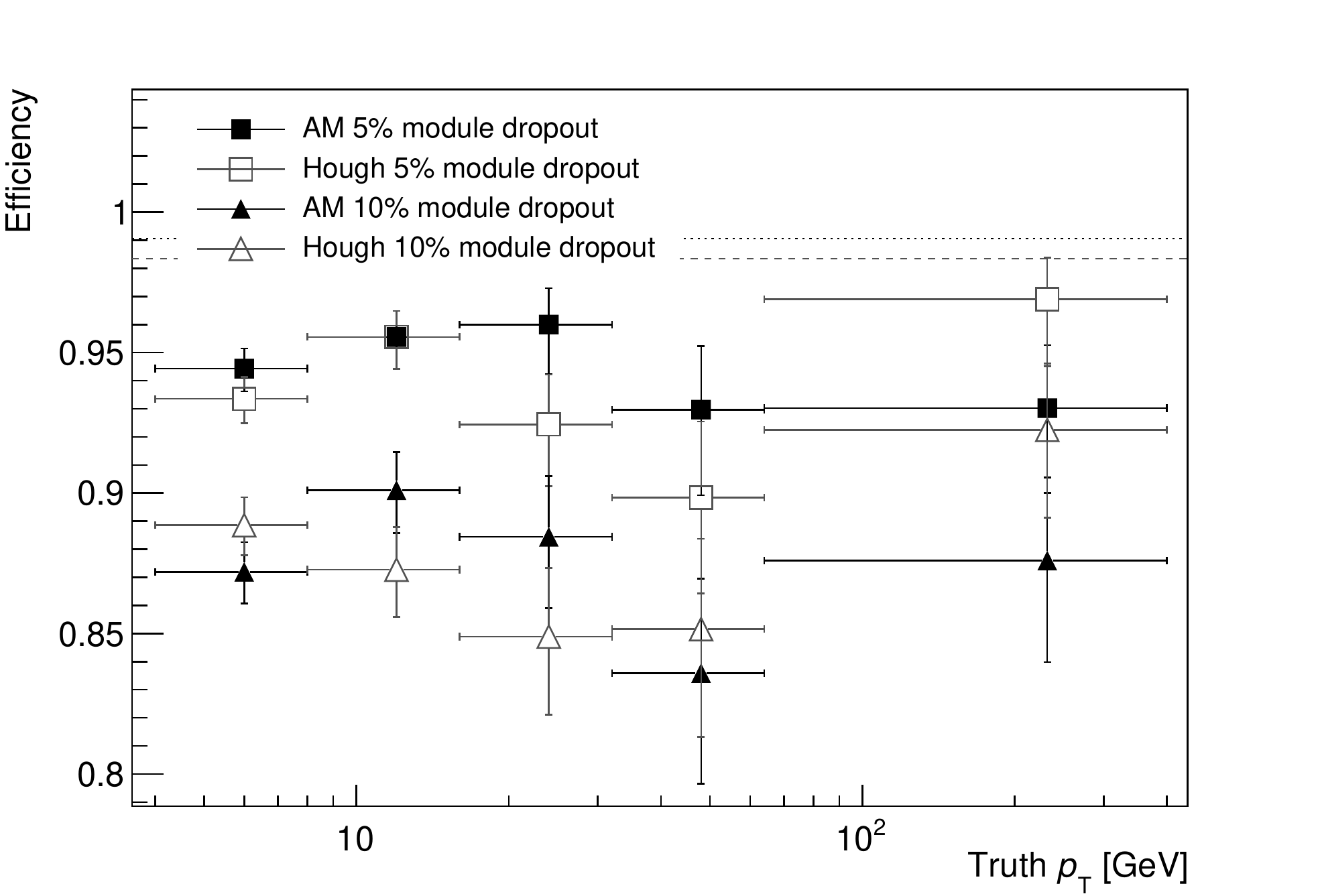}
\caption{\label{fig:turnon_module_dropout} Muon track finding efficiency of the AM pattern matching and the Hough transform as a function of the true muon \gls{pt} for single muons embedded in 200 minimum bias events when randomly dropping \SI{5}{\%} and \SI{10}{\%} of the detector modules. For reference, the dotted and dashed lines show the average nominal efficiency for the AM method and Hough transform respectively.}
\end{figure}

The muon track finding efficiency as a function of the true muon \gls{pt} is shown in figure \ref{fig:turnon_module_dropout}. It exhibits some interesting features: for instance, the Hough transform shows a drop in efficiency at \gls{pt} between 16 and \SI{64}{\GeV} but a high efficiency at higher \gls{pt}. The \gls{pt} dependence of the AM efficiency is not as strong, but there is a decrease in efficiency between 32 and \SI{64}{\GeV}. Note, however, that the efficiency has high statistical uncertainties and no firm conclusions can be drawn.

\section{Summary}
\label{sec:summary}

Both methods are able to find at least six of the muon hits in each event with an efficiency at the level of 98--\SI{99}{\%}. This efficiency is not directly dependent on the amount of pile-up. 
Both systems can be tuned to have a nearly full efficiency, e.g. by using very large superstrips or accumulator bins. 
The optimization of the hit filtering must be done so that the efficiency is kept high while respecting the hardware constraints and rejecting enough hits to keep the
latency of the fitter low. 
The number of track fits, as shown in figure~\ref{fig:combinations}, grows rapidly and non-linearly with more pile-up for both methods.
With modern FPGAs capable of performing a track fit every nanosecond, both methods yield a mean number of hit combinations that could be fitted within a latency of 
a few microseconds. However, the tails of the distributions are long and both methods will likely produce some events with several thousand hit combinations, which might be truncated
by the latency limit. The AM pattern matching show a consistently lower mean and \SI{75}{\%} percentile over the range of pile-up hits tested here. Thus, this option seem to give the best
safety limit when it comes to fitting latency.

Another important factor for a track trigger is the quality of the tracks, e.g. the final trigger decision could be based on the $\chi^2$ and the \gls{pt} measurement.
Since in general the track fit quality will increase with the number of hits used, it is important for the hit filter to send as many of the true track hits as possible to the fitter
within the same pattern or group. As shown in figures~\ref{fig:muonHitsDifference} and~\ref{fig:nHitsCorrectlyAssigned}, the AM pattern matching finds more muon hits than the Hough transform
in about \SI{40}{\%} of the events and has a larger fraction of 8 out of 8 hits found.

Given these results, the AM pattern matching looks like the best option for a hardware track trigger. 
However, there are some drawbacks to this method. AM chips with specifications that meet the task are not commercially available and have to be custom-made, which is expensive. 
The Hough transform, on the other hand, can be implemented in commercially available FPGAs. 
Both methods could be made resilient for the event of severe detector failure, e.g. if a whole detector layer stops working, by keeping backup pattern banks and uploading them to
the AM chips or simply switching the layer parameters in the Hough transform. If the beam parameters change or the pile-up conditions differ from the simulations, the Hough transform
offers greater flexibility as it can run with an arbitrary number of layers which can be increased if need be, while the AM chips must be designed for a fixed number.

\appendix
\section{Detector geometry}

The detector geometry is described in general in section \ref{sec:simulation}. Table \ref{tab:BarrelGeometry} presents a detailed description of the geometric properties of the detector.

\begin{table}[htbp]
\centering
\caption{\label{tab:BarrelGeometry} The number of modules in $z$ and $\phi$, and the active area on the modules for the barrel layers. The layers with bold numbers are the ones used for the patterns and Hough transform.}
\smallskip
\begin{tabular}{|l|c|c|c|c|c|c|}
\hline
Layer & Radius [mm] & $N_z$ & $N_{\phi}$ & length [mm] & width [mm] & Type\\
\hline
0     & 35   & 8  & 12 & 40 & 20  & pixel\\
1     & 98   & 10 & 22 & 40 & 40  & pixel\\
2     & 165  & 10 & 30 & 40 & 40  & pixel\\
3     & 233  & 10 & 42 & 40 & 40  & pixel\\
\bf{4}& 300  & 10 & 52 & 40 & 40  & pixel\\
\bf{5},\bf{6}   & 400  & 60 & 28 & 24 & 100 & strip\\
\bf{7},\bf{8}   & 550  & 60 & 38 & 24 & 100 & strip\\
9,\bf{10}  & 700  & 30 & 50 & 48 & 100 & strip\\
11,\bf{12} & 850  & 30 & 60 & 48 & 100 & strip\\
13,\bf{14} & 1000 & 30 & 74 & 48 & 100 & strip\\
\hline
\end{tabular}
\end{table}

\end{document}